\documentclass[a4paper,12pt]{article}
\usepackage{graphicx}
\usepackage{psfrag}

\usepackage{mathrsfs}
\usepackage{amssymb}
\usepackage{amsfonts}
\usepackage{latexsym}
\usepackage{amsmath}
\usepackage{amsthm}
\usepackage[cp1251]{inputenc}
\usepackage[english,russian]{babel}

\newcommand{\er}[1]{\textrm{(\ref{#1})}}
\def\lb{\label}

\theoremstyle{plain}

\newtheorem{theorem}{\bf Теорема}[section]
\newtheorem{lemma}[theorem]{\bf Лемма}
\newtheorem{proposition}[theorem]{\bf Предложение}

\theoremstyle{remark}

\def\1{1\!\!1}
   \def\all{\text{для всех}}

\def\a{\alpha}    
\def\b{\beta}     
    
\def\G{\Gamma}    \def\mD{{\mathscr D}}
\def\d{\delta}    
\def\D{\Delta}    
      
     \def\mH{{\mathscr H}}
      
\def\p{\psi}

\def\l{\lambda}\def\cM{{\cal M}}

\def\r{\rho}   \def\cQ{{\cal Q}}   
\def\s{\sigma}    \def\mR{{\mathscr R}}
 \def\cS{{\cal S}}   
\def\t{\tau}

\def\o{\omega}    
       
    \def\cZ{{\cal Z}}

\def\vp{\varphi}
\def\vk{\varkappa}

\newcommand{\gS}{\mathfrak{S}}

\def\mD{{\mathscr D}}

\def\mH{{\mathscr H}}

\def\mR{{\mathscr R}}

\def\Z{{\Bbb Z}}
\def\R{{\Bbb R}}
\def\C{{\Bbb C}}
\def\T{{\Bbb T}}
\def\N{{\Bbb N}}

\def\qqq{\qquad}
\def\qq{\quad}

\let\ge\geqslant
\let\le\leqslant

\newcommand{\ca}{\begin{cases}}
\newcommand{\ac}{\end{cases}}
\newcommand{\ma}{\begin{pmatrix}}
\newcommand{\am}{\end{pmatrix}}

\def\lt{\biggl}
\def\rt{\biggr}

\renewcommand{\[}{\begin{equation}}
\renewcommand{\]}{\end{equation}}

\def\wt{\widetilde}
\def\pa{\partial}
\def\sm{\setminus}

\def\no{\noindent}
\def\ol{\overline}
\def\iy{\infty}

\def\/{\over}
\def\ts{\times}

\def\os{\oplus}
\def\ss{\subset}

\def\Re{\mathop{\rm Re}\nolimits}

\def\Im{\mathop{\rm Im}\nolimits}

\def\Tr{\mathop{\rm Tr}\nolimits}
\def\const{\mathop{\rm const}\nolimits}

\def\BBox{\hspace{1mm}\vrule height6pt width5.5pt depth0pt \hspace{6pt}}

\begin{document}

\title {Оператор третьего порядка с периодическими
коэффициентами на вещественной оси}

\author{Андрей Баданин
\begin{footnote}
{ Северный (Арктический) федеральный университет,
Архангельск, наб. Северной Двины, 17,
e-mail: an.badanin@gmail.com}
\end{footnote}
 \and Евгений Коротяев
\begin{footnote}
{ Санкт-Петербургский государственный университет,
Санкт-Петербург, Университетская наб., 7-9, e-mail: korotyaev@gmail.com}
\end{footnote}
}

\maketitle

УДК 517.984.5

Ключевые слова: оператор третьего порядка с периодическими коэффициентами,
спектр, асимптотики

\begin{abstract}
\no Рассматривается оператор третьего порядка с периодическими
коэффициентами на вещественной оси. Этот оператор связан с задачей
интегрирования нелинейного эволюционного уравнения Буссинеска. При
минимальных условиях на гладкость коэффициентов доказываются
следующие результаты: 1) оператор самосопряжён и раскладывается в
прямой интеграл, 2) спектр оператора абсолютно непрерывен, заполняет
всю ось и имеет кратность один  или три, 3) построена и исследована
функция Ляпунова, аналитическая на трехлистной римановой
поверхности, 4) спектр кратности три ограничен и выражен в терминах
некоторой целой функции (дискриминанта).
\end{abstract}


\section {Введение и основные результаты}
\setcounter{equation}{0}

Рассмотрим дифференциальный оператор третьего порядка
\[
\lb{Hpq}
H=i\pa^3+i p\pa+i \pa p+q
\]
действующий в $L^2(\R)$, где вещественные 1-периодические
коэффициенты $p,q$ принадлежат пространству $L^1(\T),\T=\R/\Z$, с
нормой $\|f\|_{L^1(\T)}=\int_0^1|f(s)|ds$. В предложении \ref{thH}
мы докажем, что оператор $H$ является самосопряженным на области
определения
\[
\lb{cDH} \mD(H)=\Bigl\{f\in L^2(\R):i(f''+pf)'+ipf'+qf\in L^2(\R),
f'',(f''+pf)'\in L^1_{loc}(\R)\Bigr\}.
\]

Оператор $H$ применяется в методе обратной задачи интегрирования
нелинейного эволюционного уравнения Буссинеска на окружности (“bad
Boussinesq” см. \cite{McK}):
\[
\lb{Be} \ddot p=\pa^2\Bigl({4\/3}p^2+{1\/3}\pa^2 p\Bigr),\qq\dot
p=\pa q.
\]
 Именно,  уравнениe Буссинеска \er{Be}  равносильно
 нелинейному уравнению Лакса $\dot H=HK-KH$, где
$K=-\pa^2+{4\/3}p$. Здесь $ \dot f$ и $\pa f$ обозначает производную
функции $f$ по  временной  и, соответственно, пространственной
переменной.

Известно, что если $p,q\in C^\iy(\T)$, то самосопряженный оператор
$H$ может быть определен как замыкание соответствующего минимального
оператора. Более того, спектр $\s(H)$ оператора $H$ абсолютно непрерывен (см.,
напр., \cite{DS}, гл.~ XIII) и
заполняет всю ось (см.\cite{McG}).
Наша цель -- определить самосопряженный
оператор $H$ для более широкого класса коэффициентов $p,q\in
L^1(\T)$ и описать спектр в терминах так называемой функции
Ляпунова. Такое описание удобно для анализа спектра (см., напр.,
\cite{BK2}, \cite{CK1},  \cite{K1}, \cite{K2}).

Огромное число статей посвящено прямым и обратным спектральным
задачам для оператора
Шрёдингера с периодическим потенциалом: Дубровин \cite{D},
Гарнетт -- Трубовиц \cite{GT}, Итс -- Матвеев \cite{IM},
Каргаев -- Коротяев \cite{KK}, Марченко -- Островский
\cite{MO} и т.д.  Заметим, что Коротяев \cite{K3} распространил
результаты \cite{MO}, \cite{GT},\cite{KK}
для случая $-y''+qy$  на случай периодических распределений,
т.е. $-y''+q'y$ на $L^2(\R)$, где периодический $q\in
L_{loc}^2(\R)$.

Кратко опишем результаты для векторных дифференциальных уравнений.
Обратная задача (включая характеризацию) для векторнозначного
оператора Штурма -- Лиувилля на конечном интервале с условиями
Дирихле была решена недавно в работах Коротяева и Челкака
\cite{CK2},  \cite{CK3}. Периодический случай сложнее и многие
работы посвящены только прямой спектральной задаче для периодических
систем: Карлсон \cite{Ca}, Гельфанд -- Лидский
\cite{GL}, Гестези и соавторы \cite{CHGL}, Коротяев и
Челкак \cite{CK1}, Крейн \cite{Kr} и т.д. Опишем важные для нас в дальнейшем
результаты из
\cite{BBK}, \cite{CK1}, \cite{K1}, \cite{K2} для операторов первого и
второго порядка с периодическими матричнозначными потенциалами:

1) построена и изучена функция Ляпунова,  аналитическая на римановой
поверхности,

2) построено конформное отображение с вещественной частью,
заданной интегрированной плотностью состояний,
и мнимой частью, заданной показателем Ляпунова,
и изучены его основные свойства,

3) получены формулы следов (аналогичные скалярному случаю),

4) получены оценки длин лакун в терминах потенциала,

5) получены асимптотики собственных значений периодической и антипериодической задач
и точек ветвления функции Ляпунова,

6) показано, что края лакун в спектре оператора являются
периодическими или антипериодическими
соб\-ст\-вен\-ны\-ми
значениями или точками ветв\-ле\-ния функ\-ции Ля\-пу\-но\-ва.

Спектральный анализ операторов высокого ($\ge 3$) порядка с
периодическими коэффициентами сильно усложняется тем
обстоятельством, что матрица монодромии содержит как элементы,
которые ограничены при больших вещественных значениях спектрального
параметра, так и растущие элементы. Напомним также, что функция
Ляпунова для оператора второго порядка является целой, а для
оператора $2p$-го порядка -- $p$-листной, см. \cite{BK3}. Конформное
отображение, важное в спектральном анализе операторов с
периодическими коэффициентами, для операторов высокого порядка до
сих пор не построено. Операторы четного ($\ge 4$) порядка с
периодическими коэффициентами рассматривались в работах: Баданин --
Коротяев \cite{BK1}, \cite{BK2}, \cite{BK3}, Папаниколау \cite{P1},
\cite{P2}, Ткаченко \cite{Tk}, см. также ссылки в этих работах.

Спектральные свойства периодического уравнения Эйлера-Бернулли
$(ay'')''=\l by$ изучались Папаниколау в \cite{P1}, \cite{P2}.
Показано, что спектр лежит на положительной полуоси и
является объединением неперекрывающихся зон кратности 2,
аналогично случаю оператора Хилла. Начало спектра является одновременно
простым периодическим собственным значением и точкой ветвления функции Ляпунова.
Все другие ветвления лежат на отрицательной полуоси.

В работе \cite{BK3}  получены следующие результаты об операторе
$2p$-го ($p\ge 2$) порядка с периодическими коэффициентами (случай
$p=2$ см. в \cite{BK1}, \cite{BK2}): построена и изучена функция
Ляпунова, аналитическая на $p$-листной римановой поверхности,
получены асимптотики собственных значений периодической и
антипериодической задач и точек ветвления функции Ляпунова.
Края лакун в спектре такого оператора являются
периодическими или антипериодическими
соб\-ст\-вен\-ны\-ми
значениями или точками ветв\-ле\-ния функ\-ции Ля\-пу\-но\-ва,
а кратность спектра может быть равна любому четному числу от
2 до $2p$.
При больших энергиях края лакун являются
периодическими или антипериодическими соб\-ст\-вен\-ны\-ми
значениями и спектр  имеет кратность 2.

Гораздо менее изучен оператор нечетного порядка с негладкими
периодическими коэффициентами.
В сущности, о спектре такого оператора
до настоящего времени ничего не известно.

Прямая и обратная задача рассеяния для оператора третьего порядка с
убывающими коэффициентами рассмотрена  в работе Дейфта--Томеи--Трубовица
\cite{DTT} (там же см. ссылки), где, в частности, достаточно хорошо
развита спектральная теория. Дальнейшие исследования в этом
направлении изложены в книге Билза--Дейфта--Томеи \cite{BDT} и в работе
Суханова \cite{Su}. Отметим также работы Амура \cite{A1}, \cite{A2},
где рассматривался оператор третьего порядка на конечном интервале с
краевыми условиями, представляющими из себя комбинацию
квазипериодических условий и условий Дирихле.
Несамосопряженный оператор третьего
порядка с гладкими периодическими коэффициентами рассматривался в
работе МакКина \cite{McK} в связи с задачей интегрирования так
называемого уравнения “good Boussinesq” на окружности.

В данной работе мы начинаем систематическое исследование спектра
самосопряженного оператора $H$ с периодическими коэффициентами из
класса $L^1(\T)$. Изучаемый нами оператор связан с задачей
интегрирования уравнения “bad Boussinesq” \er{Be}, см. \cite{McK}.
Следуя схеме \cite{K2}, мы вводим функцию Ляпунова $\D(\l)$,
аналитическую на трехлистной римановой поверхности и удовлетворяющую
обычному равенству $\D=\cos k(\l)$, где $k$ -- квазиимпульс.
Трехлистность функции Ляпунова существенно усложняет спектральный
анализ оператора $H$.
Далее, используя
разложение оператора $H$ в прямой интеграл операторов, действующих
на конечном интервале, мы доказываем, что спектр оператора $H$
абсолютно непрерывный,  и описываем его в терминах функции Ляпунова.
При этом мы показываем, что значения функции Ляпунова на
вещественной оси определяют спектр таким же образом, как и в случае
оператора Хилла. Края спектральных интервалов со спектром
кратности 3 являются точками ветвления функции Ляпунова.
Подчеркнем, что для операторов третьего порядка
(в отличие от операторов четного порядка)
периодические и антипериодические собственные значения никак не
связаны с кратностью спектра.

Полученные здесь результаты мы используем в работах \cite{BK4},
\cite{BK5}. В работе \cite{BK4} изучается случай  малых $p,q\to 0$.
Мы доказываем, что в этом случае весь спектр оператора $H$ имеет
кратность 1, за возможным исключением одного маленького интервала со
спектром кратности 3 в окрестности нуля. Получены явные условия на
$p,q$, при которых такого интервала нет и условия, при которых он есть,
при этом
найдена его асимптотика. В работе \cite{BK5} мы исследуем риманову
поверхность функции Ляпунова, находим асимптотики ее точек ветвления
и асимптотики собственных значений периодической и
антипериодической задач для уравнения $i y'''+ipy'+i(py)'+q y=\l y$
при высоких энергиях. Показано, что в случае "общих" коэффициентов
$p,q$ риманова поверхность функции Ляпунова имеет бесконечный род.

Для разложения оператора $H$ в прямой интеграл введем гильбертовы
пространства
\[
\lb{intH} \mH'=L^2([0,1],dt),\qqq \mH=\int_{[0,2\pi)}^\os \mH' \ {d
k\/2\pi}
\]
и операторы
$$
H( k)=i\pa^3+i\pa p+ip\pa +q,\qq  k\in[0,2\pi),
$$
действующие в $\mH'=L^2(0,1)$ и самосопряженные на области
определения
\begin{multline}
\lb{cDA}
\mD(H( k))=\Bigl\{f\in L^2(0,1):i(f''+pf)'+ipf'+qf\in L^2(0,1),f'',(f''+pf)'\in L^1(0,1),\\
f_j(1)=e^{i k}f_j(0)\ \all\ j=1,2,3,\ \text{где}\
f_1=f,f_2=f',f_3=f''+pf\Bigr\},
\end{multline}
см. лемму \ref{PrA}. Введем унитарный  оператор $U:L^2(\R)\to\mH$
равенством
\[
\lb{gt} (U f)_k(t)=\sum_{n\in\Z}e^{-in k}f(t+n),\qqq
(k,t)\in[0,2\pi)\ts[0,1].
\]
  Приведем предварительный результат о разложении оператора $H$ в
прямой интеграл.

\begin{proposition}
\lb{thH}
Оператор $H$, определенный в \er{Hpq},
 \er{cDH}, является самосопряженным и удовлетворяет
равенству
\[
\lb{deH} UHU^{-1}=\int_{[0,2\pi)}^\os H( k){d k\/2\pi}.
\]
\end{proposition}

\no {\bf Замечания.}
1) Аналогичный результат для оператора второго
порядка хорошо известен, см. \cite{Ge}, \cite{RS}. Для оператора
произвольного четного порядка с гладкими коэффициентами
соответствующие результаты см. в статье Ткаченко \cite{Tk}.

2) В нашем случае оператор имеет коэффициенты из
$L^1(0,1)$ и поэтому его область определения более сложная,
чем у оператора с гладкими коэффициентами.
Кроме того, оператор имеет нечетный порядок
и неполуограничен снизу. Этот случай требует отдельного анализа.

В случае когда коэффициенты $p,p',q\in L^1(\T)$, можно определить
стандартную матрицу монодромии
$$
\wt M(\l)
=\{\wt\vp_{j}^{(k-1)}(1,\l)\}_{k,j=1}^3,\qqq\l\in\C,
$$
где $\wt\vp_{j}$ -- решения уравнения
\[
\lb{1b} i y'''+ipy'+i(py)'+q y=\l y,\qq (t,\l)\in\R\ts\C,
\]
удовлетворяющие начальным условиям
$\wt\vp_{j}^{(k-1)}(0,\l)=\d_{jk},j,k=1,2,3$
(см., напр., \cite{DS}, гл.~XIII.7).
Если коэффициент  $p\in L^1(\T)$ и $p'\notin L^1(\T)$, то
стандартная матрица монодромии уже  не определена, поскольку, вообще
говоря, производная $y''$ не является непрерывной. Здесь требуется
существенная модификация. Мы введём модифицированную матрицу
монодромии $ M(1,\l)$ по формуле
\[
\lb{deM}
 M(t,\l)=\ma\vp_1&\vp_2&\vp_3\\
\vp_1'&\vp_2'&\vp_3'\\
\vp_1''+p\vp_1&\vp_2''+p\vp_2&\vp_3''+p\vp_3\am(t,\l),\qq
(t,\l)\in\R\ts\C.
\]
Здесь $\vp_1,\vp_2, \vp_3$ есть фундаментальные
решения уравнения \er{1b},
удовлетворяющие ус\-ловиям
\[
\lb{ic}  M(0,\l) =\1_3,\qqq\l\in\C,
\]
где $\1_N$ -- единичная $N\ts N$ матрица.
Соответствующий
характеристический полином $D$ имеет вид
\[
\lb{1c} D(\t,\l)=\det( M(1,\l)-\t \1_{3}),\qq (\t,\l)\in\C^2.
\]
Собственное значение матрицы $ M(1,\l)$ называется {\it
мультипликатором}, оно является нулем алгебраического уравнения
$D(\cdot,\l)=0$. Каждая $3\ts 3$-матрица $ M(1,\l), \l\in\C$, имеет
ровно $3$ (с учетом кратности) мультипликатора $\t_j(\l),j=1,2,3$. В
случае $p=q=0$ мультипликаторы имеют вид
$$
\t_j^0(\l)=e^{i\o^{j-1}z},\qqq \l\in \C,\qq j=1,2,3,\qqq \text{где} \qqq
 \o=e^{i{2\pi\/3}},
$$
здесь и далее мы считаем
$$
z=\l^{1\/3},\qqq \arg\l\in\Bigl(-{\pi\/2},{3\pi\/2}\Bigr],\qqq
\arg z\in\Bigl(-{\pi\/6},{\pi\/2}\Bigr].
$$
Введем функцию
$$
T(\l)=\Tr M(1,\l), \qqq \l\in \C.
$$

\begin{theorem} \lb{TMM}
i) Матричнозначная функция $ M(1,\cdot)$ является целой и для всех
$\t,\l\in\C$ верны равенства:
\[
\lb{idM}  M^*(1,\ol\l)J M(1,\l)=J, \qq\text{где}\qq J=\ma
0&0&i\\
0&-i&0\\
i&0&0 \am,
\]
\[
\lb{cM2} D(\t,\l)=\det( M(1,\l)-\t \1_{3})=-\t^3+\t^2T(\l)-\t \ol T(\ol\l)+1,
\]
\[
\lb{cM} \det M(1,\l)=1.
\]

ii) Пусть $\l\in\R$. Если $\t(\l)$ является мультипликатором, то
$\bar\t^{-1}(\l)$ также является мультипликатором.
Возможны только два случая:

 a) все три мультипликатора лежат на единичной окружности;

 b) ровно один (простой) мультипликатор лежит на единичной
окружности.

\no Более того, в случае b) мультипликаторы имеют вид
\[
\lb{mrs}
e^{i k},\qq
e^{i\bar k}, \qq e^{-i2\Re k}\qq\text{для некоторого}
\qq   k\in\C:\Im k\ne 0.
\]

iii) Пусть $p,p',q\in L^1(\T)$. Тогда модифицированная матрица
монодромии $M$ и стандартная матрица монодромии $\wt M$ связаны
равенством
\[
\lb{MwM} M(1,\cdot)=\cS^{-1}\wt M(\cdot)\cS,\qq \qqq
\text{где}\qqq \cS=\ma1&0&0\\0&1&0\\-p(0)&0&1\am.
\]

\end{theorem}

{\bf Замечание} 1) Равенство \er{MwM} показывает, что в случае
гладких коэффициентов $p,q$ собственные значения матриц $M(1,\l)$
и $\wt M(\l)$ совпадают, т.е. мультипликаторы могут определяться,
как собственные значения любой из этих матриц.

2) Мы считаем, что даже в случае гладких коэффициентов
модифицированная матрица $M$ более удобна для анализа оператора
$H$, чем матрица $\wt M$. В частности, соотношение симплектичности
\er{idM} для матрицы $M$ имеет более простой вид, чем аналогичное
равенство для $\wt M$.

Коэффициенты полинома $D$ -- целые функции переменной $\l$. Известно
(см., напр., \cite{Fo}, гл.~8), что его корни $\t_j(\l),j=1,2,3$,
составляют одну или несколько ветвей одной или нескольких
аналитических функций, имеющих только алгебраические особенности в
$\C$. В теореме \ref{T1r} мы показываем, что $\t_j(\l),j=1,2,3$, все
различны и составляют три ветви одной функции $\t(\l)$,
аналитической на некоторой связной трехлистной римановой поверхности
$\mR$.
Введем  {\it функции Ляпунова}
$$
\D_j={1\/2}(\t_j+\t_j^{-1}),\qqq j=1,2,3.
$$
Эти функции являются ветвями функции $\D$, аналитической на
поверхности $\mR$.  Введем {\it дискриминант} $\r(\l),\l\in\C$,
полинома $D(\cdot,\l)$ (см. \er{cM2}) равенством
\[
\lb{r}
\r=(\t_1-\t_2)^2(\t_1-\t_3)^2(\t_2-\t_3)^2.
\]
Если $p=q=0$, то функция Ляпунова $\D^0$, ее ветви $\D_j^0$ и
дискриминант $\r^0$ имеют вид
$$
\D^0=\cos\l^{1\/3},\qq\D_j^0=\cos z\o^{j-1},\qq j=1,2,3,
$$
\[
\lb{ro0} \r^0=64\sinh^2{\sqrt 3 z\/2}\sinh^2{\sqrt 3 \o z\/2}
\sinh^2{\sqrt 3 \o^2z\/2}.
\]
Мы доказываем следующие
результаты.

\begin{theorem}  \lb{T1r}
i) Функции $\t_j, j=1,2,3$, являются ветвями некоторой функции
$\t$, аналитической на связной трехлистной римановой
поверхности, и  удовлетворяют асимптотике
 \[
\lb{awt}
\t_j(\l)=e^{iz\o^{j-1}}(1+O(z^{-1}))\qqq \text{при} \qq |\l|\to\iy,
\qqq \text{где} \qq
\o=e^{i{2\pi\/3}}.
\]
ii) Функции $\D_j, j=1,2,3$, являются ветвями некоторой  функции
$\D$, аналитической на связной трехлистной римановой
поверхности, и  удовлетворяют асимптотике
\[
\lb{aD} \D_j(\l)=\cos (z\o^{j-1})+O\lt({e^{|\Im
(z\o^{j-1})|}\/|z|}\rt)\qqq \text{при} \qqq |\l|\to\iy.
\]
iii) Функция $\r$ - целая, вещественная на $\R$, и при всех $\l\in\R$ удовлетворяет равенству
\[
\lb{rtr} \r(\l)= |T(\l)|^4-8\Re T^3(\l)+18|T(\l)|^2-27.
\]

\end{theorem}

\no {\bf Замечания.}
1) Асимптотика матрицы монодромии для оператора второго порядка выражается через $\cos\sqrt{\l} $
и  $\sin\sqrt{\l} $, ограниченные на вещественной оси.
Асимптотика матрицы монодромии для оператора третьего порядка
выражается через
$e^{i\o^j\l^{1\/3}},j=0,1,2$, см. \er{3if},  неограниченные на вещественной оси.
Это порождает определенные трудности при вычислении спектральных асимптотик.
Эти трудности удается преодолеть, используя симплектичность
матрицы монодромии, см. \er{idM}, и следующие из нее симметрии мультипликаторов,
см. теорему \ref{TMM} ii).

2) Функция $\r$ определяет точки ветвления функции Ляпунова.
Аналогичная функция для уравнения “good Boussinesq” анализируется
МакКином \cite{McK}. В частности, им получено равенство, аналогичное \er{rtr}.

В следующей теореме мы описываем спектр оператора $H$ в терминах
мультипликаторов и функции Ляпунова.

\begin{theorem}
\lb{spec}
i) Спектр $\s(H)$ оператора $H$ абсолютно непрерывен и равен
\begin{multline}
\lb{sdD}
\s(H)=\{\l\in\R:|\t_j(\l)|=1\ \text{для}\ j=1,2\ \text{или}\ 3\}\\
=\{\l\in\R:\D_j(\l)\in[-1,1]\ \text{для}\ j=1,2\ \text{или}\ 3\}.
\end{multline}
Более того, кратность спектра равна числу ветвей функции $\t(\l)$
(или $\D(\l)$), удовлетворяющих условию \er{sdD}.

ii) Спектр $\s(H)$ заполняет всю вещественную ось и имеет кратность
1 или 3:  спектр кратности 3 совпадает с ограниченным  множеством
\[
\lb{sro}
\gS_3= \{\l\in\R:\r(\l)\le 0\},
\]
и спектр имеет кратность 1 вне этого множества.

iii) Если $\D_j(\l)\in(-1,1)$ для какого-либо $(j,\l)\in\{1,2,3\}\ts\R$
и $\l$ не является точкой ветвления функции $\D_j$,
то $\D_j'(\l)\ne 0$.
\end{theorem}

\no {\bf Замечания.} 1) Спектр самосопряженного
оператора любого нечетного порядка
с гладкими периодическими коэффициентами
заполняет всю ось, см., напр.,
\cite{McG}.

2) Доказательство абсолютной непрерывности спектра -- стандартное
(см., напр., \cite{RS}) и основывается на равенстве \er{deH} и том факте, что
собственные значения оператора $H(k)$ -- кусочно-аналитические и непостоянные
функции переменной $k\in[0,2\pi)$.

3) Равенство \er{sro} показывает, что кратность спектра
оператора $H$ полностью определяется значениями целой функции
$\r$.

4) Края спектральных интервалов кратности 3 являются точками ветвления
функции Ляпунова. Периодические и антипериодические собственные
значения никак не связаны с кратностью спектра.

5) Асимптотика \er{aD} и равенство \er{sdD} показывают,
что при больших энергиях ровно одна ветвь функции
Ляпунова вносит вклад в спектр, две другие ветви
принимают невещественные значения.

Приведем краткое описание работы. В $\S$ 2 мы изучаем свойства
матрицы монодромии. В $\S$ 3 мы доказываем теоремы \ref{TMM} и
\ref{T1r}. $\S$ 4 посвящен описанию оператора $H(k)$. Мы доказываем,
что этот оператор аналитически зависит от $k$ на $[0,2\pi)$ и при
каждом $k$ является самосопряженным и имеет полный набор
нормированных собственных функций. В $\S$ 5 мы изучаем оператор $H$
и доказываем предложение \ref{thH} и теорему \ref{spec}.

\section {Матрица монодромии}
\setcounter{equation}{0}

В этом параграфе мы изучим матрицу монодромии.
Мы перепишем уравнение \er{1b} в векторной форме
\[
\lb{me} Y'-P(\l)Y=Q(t) Y,\qqq (t,\l)\in\R\ts\C,
\]
где вектор-функция $Y$ и $3\ts 3$-матричнозначные функции $P, Q$
имеют вид
\[
\lb{Y} Y=\ma y_1\\y_2\\ y_{3}\am =\ma y\\y'\\y''+py\am,\qq P=\ma
0&1&0\\0&0&1\\-i\l&0&0\am,\qq Q=\ma 0&0&0\\-p&0&0\\iq&-p&0\am.
\]
При $\l\ne 0$ верно равенство
\[
\lb{PU} P=(\cZ U)(iz B)(\cZ U)^{-1},
\]
где
\[
\lb{UBZ} U={1\/\sqrt
3}\ma1&1&1\\1&\o&\o^2\\1&\o^2&\o\am=(U^*)^{-1},\qq
B=\ma1&0&0\\0&\o&0\\0&0&\o^2\am,\qq
\cZ=\ma1&0&0\\0&iz&0\\0&0&(iz)^2\am,
\]
\[
\lb{ozl} \o=e^{i{2\pi\/3}},\qq z=x+iy=\l^{1\/3},\qq
\arg\l\in\Bigl(-{\pi\/2},{3\pi\/2}\Bigr], \qq\arg
z\in\Bigl(-{\pi\/6},{\pi\/2}\Bigr].
\]
Заметим, что $3\ts 3$-матричнозначная функция $ M(t,\l)$, заданная
\er{deM}, является решением начальной задачи
\[
\lb{eM}
 M'-P(\l) M=Q(t)  M,\qqq  M(0,\l)=\1_{3}.
\]

В невозмущенном случае $p=q=0$ решение $ M_0$ уравнения \er{eM}
имеет вид $M_0=e^{t P(\l)}$. Функция $M_0(t,\l)$ является целой по
$\l$ при каждом $t\in \R$. Собственные значения матрицы $P$ равны
$iz\o^j,j=0,1,2$, и матрица $ M_0(t,\l)$ имеет собственные значения
$e^{iz\o^jt},j=0,1,2$. Из условия \er{ozl} получаем $ x\ge
\max\{0,-y\sqrt3\}$. Тогда из
$$
\Re(iz)=-y,\qqq \Re(iz\o)={y-\sqrt3 x\/2},\qqq \Re(iz\o^2)={y+\sqrt3
x\/2}
$$
следует, что для всех $\l\in\C$
\[
\lb{mO}
\max\{\Re(iz),\Re(iz\o)\}\le z_0=\Re(iz\o^2).
\]
Оценки $|e^{iz\o^jt}|\le e^{z_0|t|}$ дают
\[
\lb{eu1}
| M_0(t,\l)|\le e^{z_0|t|}\qq\text{для всех}\qq
(t,\l)\in\R\ts\C.
\]
Здесь и далее для матрицы $A$ мы используем следующую {\it норму}
$$
|A|=\max\{\sqrt{h}:h\ \text{-- собственное
значение матрицы}\ A^*A\}.
$$

Применяя преобразование подобия \er{PU} к обеим частям
равенства \er{eM}, мы получаем
\[
\lb{ecM}
\cM'-izB\cM=\cQ(t,\l)\cM,\qqq  \cM(0,\l)=\1_{3},
\]
где
\[
\lb{cMQ}
\cM=(\cZ U)^{-1}M(\cZ U),\qq
\cQ=(\cZ U)^{-1}Q(\cZ U)={1\/iz}\ U^{-1}\ma
0&0&0\\-p&0&0\\{q\/z}&-p&0\am U.
\]

Напомним, что $T=\Tr\cM(1,\cdot)$ и пусть $T_0=\Tr\cM_0(1,\cdot)$ в
невозмущенном случае.

\begin{lemma}
\lb{l21} Матричнозначная функция $
M(1,\cdot)$ является целой.
Верны неравенства
\[
\lb{esT} |T(\l)|\le 3e^{z_0+\vk}\qq  \all\qq  \l\in\C,
\]
\[
\lb{3if}
|\cM(1,\l)-e^{izB}|\le {\vk\/|z|}e^{z_0+\vk},
\]
\[
\lb{4if} |T(\l)-T_0(\l)|\le {3\vk\/|z|}e^{z_0+\vk}
\]
для всех $|\l|\ge 1$, где $\vk=\int_0^1(|p(t)|+|q(t)|)dt$ и $z_0=\Re(iz\o^2)$.
\end{lemma}

\no {\bf Доказательство.} Стандартные рассуждения,
примененные к уравнению \er{eM} показывают, что
функция $ M(t,\l)$ удовлетворяет интегральному уравнению
\[
\lb{iev}
 M(t,\l)= M_0(t,\l)+\int_0^t M_0(t-s,\l)Q(s) M(s,\l)ds.
\]
Итерации в уравнении \er{iev} дают
\[
\lb{evj}
 M(t,\l)=\sum_{n\ge 0} M_n(t,\l),\qqq  M_n(t,\l)=\int_0^t M_0(t-s,\l)Q(s) M_{n-1}(s,\l)ds.
\]
Из равенств \er{evj} следует
\[
\lb{2ig}  M_n(t,\l)=\int\limits_{0< t_1<...<
t_n<t_{n+1}=t}\prod\limits_{k=1}^{n} \lt(
M_0(t_{k+1}-t_k,\l)Q(t_k)\rt) M_0(t_1,\l)dt_1dt_2...dt_n,
\]
$ (t,\l)\in\R_+\ts\C,$ множители в произведении упорядочены справа
налево. Подставляя оценки \er{eu1} в равенства \er{2ig}, мы получаем
\[
\lb{eM1} | M_n(t,\l)|\le{e^{z_0t}\/n!}\lt(\int_0^t|Q(s)|ds\rt)^n
\qq \text{для всех} \qq (n,t,\l)\in\N\ts\R_+\ts\C.
\]
Эти оценки показывают, что
формальный ряд \er{evj} сходится абсолютно и равномерно на
любом ограниченном подмножестве в $\R_+\ts\C$.
Каждое слагаемое этого ряда является целой функцией переменной $\l$.
Сумма также обладает этим свойством.
Неравенства \er{eM1} и $|Q|\le |p|+|q|$ дают
$$
|\Tr M_n(1,\l)|\le 3| M_n(1,\l)| \le{3\vk^n\/n!}e^{z_0}\qq \text{для всех} \qq
(n,t,\l)\in\N\ts\R_+\ts\C.
$$
Отсюда имеем $ |\Tr M(1,\l)|=|\sum_0^\iy\Tr M_n(1,\l)| \le
3e^{z_0}\sum_0^\iy{\vk^n\/n!},$ что дает \er{esT}.

Решения $ \cM(t,\l)$ уравнения \er{ecM} удовлетворяет интегральному уравнению
$$
 \cM(t,\l)=e^{iztB}+\int_0^t e^{iz(t-s)B}\cQ(s)\cM(s,\l)ds,
$$
откуда следует
\[
\lb{evjc}
 \cM(t,\l)=\sum_{n\ge 0} \cM_n(t,\l),\qqq
 \cM_n(t,\l)=\int_0^t e^{iz(t-s)B}\cQ(s) \cM_{n-1}(s,\l)ds.
\]
Из равенств \er{evjc} получаем
$$
\cM_n(t,\l)=\int\limits_{0< t_1<...<
t_n<t_{n+1}=t}\prod\limits_{k=1}^{n}\lt(
e^{iz(t_{k+1}-t_k)B}\cQ(t_k)\rt) e^{izt_1B}dt_1dt_2...dt_n,
$$
что дает
$$
|\cM_n(t,\l)|\le{e^{z_0t}\/n!}\lt(\int_0^t|\cQ(s)|ds\rt)^n
\qq \text{для всех} \qq (n,t,\l)\in\N\ts\R_+\ts\C.
$$
Эти оценки показывают, что
формальный ряд \er{evjc} сходится абсолютно и равномерно на
любом ограниченном подмножестве в $\R_+\ts\C$.
Суммируя мажоранты, и используя оценку
$\int_0^1|\cQ|ds\le{\vk\/|z|}$ для всех $|z|\ge 1$
(см. \er{cMQ}),
мы получаем неравенства \er{3if}, \er{4if}.
$\BBox$

\section {Доказательство теорем \ref{TMM}-\ref{T1r}}
\setcounter{equation}{0}

\no {\bf Доказательство теоремы \ref{TMM}.} i) Из \er{eM} следует $J
M'=V M$, где
$$
V=J(P+Q)=i\ma0&0&1\\0&-1&0\\1&0&0\am\ma
0&1&0\\-p&0&1\\-i\l+iq&-p&0\am =\ma \l-q&-ip&0\\ip&0&-i\\0&i&0\am.
$$
 и $J$ определена равенством \er{idM}. Тогда $-( M^*)'J= M^*V$ для
$\l\in\R$ и
$$
( M^*J M)'=( M^*)'J M+ M^*J M' =- M^*V M+ M^*V M=0,
$$
откуда следует $( M^*J M)(t,\l)=( M^*J M)(0,\l)=J$ для всех
$(t,\l)\in\R\ts\C$, что дает \er{idM}.
Равенство \er{deM} и уравнение \er{1b} дают
$$
(\det M)'=\det \ma\vp_1&\vp_2&\vp_3\\
\vp_1'&\vp_2'&\vp_3'\\
(\vp_1''+p\vp_1)'&(\vp_2''+p\vp_2)'&(\vp_3''+p\vp_3)'\am=0.
$$
 Тогда $\det M(t,\l)=\det M(0,\l)=1$ для всех
$(t,\l)\in\R\ts\C$, что дает \er{cM}. Прямые вычисления показывают
$$
D(\t,\l)=\det ( M(1,\l)-\t \1_3)=-\t^3+\t^2\Tr M(1,\l)+B(\l)\t-1
\qq\text{для всех}\qq(\t,\l)\in\C^2,
$$
где $B(\l)=\pa_\t D(0,\l)$. Известная формула из теории матриц
(см., напр., \cite{GK}, равенство IV.1.3) дает
$$
\pa_\t D(\t,\l)=-D(\t,\l)\Tr( M(1,\l)-\t\1_3)^{-1}.
$$
Используя равенство $D(0,\l)=1$, мы получаем $B(\l)=-\Tr
M^{-1}(1,\l)$. Равенство \er{idM} дает $ M^{-1}(1,\l)=-J
M^*(1,\ol\l)J$, откуда следует $\Tr M^{-1}(1,\l)=\Tr M^*(1,\ol\l)$
для всех $\l\in\C$. Тогда $B(\l)=-\Tr M^*(1,\ol\l)$, что дает
\er{cM2}.

ii) Из равенства \er{cM2} следует
\[
\lb{dM}
D(\t,\l)=-\t^3 { \ol D(\bar\t^{-1},\l)}\qq\text{для всех}\qq
(\t,\l)\in\C\ts\R,\qq \t\ne 0.
\]
Таким образом если $\t(\l)$ является корнем $D(\t,\l)$ для
некоторого $\l\in\R$, то $\bar\t^{-1}(\l)$ также является корнем.
Используя равенство $\t_1\t_2\t_3=\det M(1,\cdot)=1$, мы получаем
нужные утверждения.

iii) Для уравнения \er{1b} введем фундаментальную матрицу
$$
\wt\cM(t,\l)=(\wt\vp_{j}^{(k-1)}(t,\l))_{k,j=1}^3,\qqq
(t,\l)\in\R\ts\C.
$$
Тогда $\wt M(\l)=\wt\cM(1,\l)$.
Поскольку $\wt\vp_j$ -- фундаментальные решения уравнения \er{1b}
и $\vp_j$ также являются его решениями, $\vp_j$ --
линейные комбинации $\wt\vp_j$.
Начальные условия \er{ic} дают
$$
\vp_1(t,\l)=\wt\vp_1(t,\l)-p(0)\wt\vp_3(t,\l),\qq
\vp_2(t,\l)=\wt\vp_2(t,\l),\qq \vp_3(t,\l)=\wt\vp_3(t,\l),
$$
для всех $(t,\l)\in\R\ts\C$, откуда следует, что
\[
\lb{M1}
\ma\vp_1&\vp_2&\vp_3\\
\vp_1'&\vp_2'&\vp_3'\\
\vp_1''&\vp_2''&\vp_3''\am(t,\l)
=\ma\wt\vp_1-p(0)\wt\vp_3&\wt\vp_2&\wt\vp_3\\
\wt\vp_1'-p(0)\wt\vp_3'&\wt\vp_2'&\wt\vp_3'\\
\wt\vp_1''-p(0)\wt\vp_3'&\wt\vp_2''&\wt\vp_3''\am(t,\l)
=\wt\cM(t,\l)\cS,\qq(t,\l)\in\R\ts\C.
\]
Поскольку $\cS^{-1}=\ma1&0&0\\0&1&0\\p(0)&0&1\am$,
равенство \er{deM} дает
\[
\lb{M2}
M(t,\l)=\cS^{-1}\ma\vp_1&\vp_2&\vp_3\\
\vp_1'&\vp_2'&\vp_3'\\
\vp_1''&\vp_2''&\vp_3''\am(t,\l)\qq\all\qq(t,\l)\in\R\ts\C.
\]
Из равенств \er{M1}, \er{M2} следует \er{MwM}.
$\BBox$

\no {\bf Доказательство теоремы \ref{T1r}.}
i)
Обозначим через $\t_j^0=e^{i\o^{j-1}z}, j=1,2,3$, собственные
значения матрицы $e^{izB}$. По известной теореме теории матриц (см.,
напр., \cite{HJ}, следствие 6.3.4) из \er{3if} следует, что при
каждом $|\l|>1$ матрица $\cM(1,\l)$, а значит и
матрица $ M(1,\l)$, имеет по крайней мере одно собственное значение
$\t_j(\l)$ в каждом круге с центром $\t_j^0(\l), j=1,2,3$, и
радиусом ${\vk\/|z|}e^{z_0+\vk}$. В частности, при $j=3$ получаем
\[
\lb{e16}
|\t_3(\l)-e^{i\o^2z}|<{\vk\/|z|}e^{z_0+\vk}\qq
\text{для всех}\ \ |\l|>1,
\]
откуда следует
$|\t_3(\l)e^{-i\o^2z}-1|<{\vk\/|z|}e^{\vk}$
для всех $|\l|>1,$
что дает асимптотику
\er{awt} для $j=3$.
Используя равенства $\t_2=\bar\t_3^{-1},\t_1=(\t_2\t_3)^{-1}$,
получаем \er{awt} для $j=1,2$.

Кроме того, асимптотика \er{awt} показывает, что все функции
$\t_j(\l),j=1,2,3$, различны. Предположим, что какая-либо из функций
$\t_j(\l)$ является целой функцией переменной $\l$. Асимптотика
\er{awt} показывает, что она является целой функцией порядка
${1\/3}$ и, следовательно, имеет бесконечное число нулей (см.,
напр., \cite{Le}, гл. I.10). Это противоречит равенству
$\t_1\t_2\t_3=1$. Таким образом, ни одна из функций $\t_j$ не
является целой и тогда $\t_j$ -- три ветви одной функции
$\t$, аналитической на связной трехлистной
римановой поверхности.

ii) Асимптотика \er{awt} дает \er{aD}.
Асимптотика \er{aD} показывает, что все функции
$\D_j(\l),j=1,2,3$, различны. Предположим, что какая-либо из функций
$\D_j$ является целой функцией.
Тогда $\t_j=\D_j+\sqrt{\D_j^2-1}$ является аналитической функцией
на двулистной римановой поверхности, что противоречит утверждению ii).
Следовательно, ни одна из функций $\D_j$ не
является целой и тогда $\D_j$ -- три ветви одной функции
$\D$, аналитической на связной трехлистной
римановой поверхности.

iii) Функция $\r$ является дискриминантом кубического полинома \er{cM}
с целыми коэффициентами, поэтому $\r$ - целая функция.
Стандартная формула для дискриминанта $d$ кубического полинома
$-\t^3+a\t^2-b\t+1$ дает $d=(ab)^2-4(a^3+b^3)+18ab-27$,
откуда следует \er{rtr}.
$\BBox$

Нам потребуются следующие свойства целых функций
$D(e^{ik},\cdot),k\in[0,2\pi)$.

\begin{lemma}
\lb{CLD}

i) Каждая из
функций $D(e^{ik},\l),k\in[0,2\pi),$ удовлетворяет асимптотике
\[
\lb{asD0}
D(e^{ik},\l)=D_0(e^{ik},\l)\bigl(1+O(|z|^{-1})\bigr)\qq\text{при}\qq |\l|\to\iy,
\]
если $|z-k-2\pi n|\ge{\pi\/2}, |z\o-k+2\pi n|\ge{\pi\/2}$ для всех
$n\in\N$, где
\[
\lb{iD0}
D_0(\t,\l)=-\t^3+\t^2T_0(\l)-\t\ol T_0(\ol\l)+1.
\]

ii) Существует $n_0\ge 1$ такое, что
для каждого целого $N>n_0$
и

\no a) для всех $k\in[0,{\pi\/2})\cup({3\pi\/2},2\pi)$ функция
$D(e^{ik},\cdot)$ имеет ровно $2N+1$ нуль, с учетом кратности,
в круге $\{\l:|\l|<(\pi(2N+1))^3\}$;

\no b) для всех  $k\in[{\pi\/2},{3\pi\/2}]$, функция
$D(e^{ik},\cdot)$ имеет ровно $2N$ нулей, с учетом кратности,
в круге $\{\l:|\l|<(2\pi N)^3\}$.

\no Более того, для каждого $n>N$ и для всех  $k\in[0,2\pi)$ функция
$D(e^{ik},\cdot)$ имеет
ровно один простой нуль в каждом круге $\{\l:|z-k-2\pi n|<{\pi\/2}\},
\{\l:|z\o-k+2\pi n|<{\pi\/2}\}$.
Других нулей эта функция не имеет.

\end{lemma}

\no {\bf Proof.}
i) Пусть $k\in[0,2\pi)$.
Из равенств \er{cM2}, \er{iD0} получаем
$$
|D(e^{ik},\l)-D_0(e^{ik},\l)|=|e^{ik}(T(\l)-T_0(\l))-\ol T(\ol\l)
+\ol T_0(\ol\l)|\qq\all\qq\l\in\C.
$$
Оценки \er{4if} дают $|T(\l)-T_0(\l)|\le{3\vk\/|z|}e^{z_0+\vk}$,
и тогда
\[
\lb{eD1}
|D(e^{ik},\l)-D_0(e^{ik},\l)|\le {6\vk\/|z|}e^{z_0+\vk}\qq\all\qq|\l|\ge 1.
\]
Предположим, что
\[
\lb{eD2}
|D_0(1,\l)|\ge {e^{z_0}\/8},\qq\Bigl\{|\l|>R:
\bigl|z-k-2\pi n\bigr|\ge{\pi\/2},\ \bigl|z\o-k+2\pi n\bigr|\ge{\pi\/2}\ \all\  n\in\N\Bigr\}
\]
для некоторого достаточно большого $R>0$.
Оценки \er{eD1} и \er{eD2} дают \er{asD0}.

Докажем \er{eD2}. Используя равенство
$D_0(\t,\l)=-(\t-e^{iz})(\t-e^{i\o z})(\t-e^{i\o^2 z})$,
мы получаем
$$
D_0(e^{ik},\l)=-(e^{ik}-e^{iz})(e^{ik}-e^{i\o z})(e^{ik}-e^{i\o^2 z})
=-i8e^{i{3k\/2}}\sin{z-k\/2}\sin{z\o-k\/2}\sin{z\o^2-k\/2}
$$
для всех $\l\in\C$.
Используя стандартную оценку
$|\sin z|>{1\/4}e^{|\Im z|}$ при $|z-\pi n|\ge{\pi\/4}$
для всех $n\in\Z$ (см. \cite{PT}, лемма 2.1) в \er{iD0}, мы получаем
$$
|D_0(e^{ik},\l)|
>{1\/8}e^{{1\/2}(|\Im z|+|\Im z\o|+|\Im z\o^2|)}\ge{e^{z_0}\/8}
$$
для всех $\{\l\in\C:|z\o^j-k-2\pi n|>1,j=0,1,2, n\in\N\}$.
Учитывая, что неравенства $|z\o^2-k\pm 2\pi n|>1,|z-k+2\pi n|\ge{\pi\/2},
|z\o-k-2\pi n|\ge{\pi\/2}, n\in\N$, выполнены для всех
достаточно больших $|\l|$, мы получаем \er{eD2}.

ii) Рассмотрим $k\in[0,{\pi\/2})$. Доказательство для
других значений $k$ аналогично.
Пусть $N\ge 1$ - достаточно большое и $N'>N$ - любое целое число.
Введем контуры $C_\a(r)=\{\l:|z-\a|=r\},r>0,\a\ge 0$.
Пусть $\l$ принадлежит контурам
$$
C_0(\pi(2N+1)),\qq
C_0(\pi(2N'+1)),
\qq C_{k+2\pi n}\Bigl({\pi\/2}\Bigr),\qq
C_{(k-2\pi n)\o^2}\Bigl({\pi\/2}\Bigr),\qq n>N.
$$
Асимптотика \er{asD0} дает
$$
|D(e^{ik},\l)-D_0(e^{ik},\l)|=|D_0(e^{ik},\l)|\Bigl|{D(e^{ik},\l)\/D_0(e^{ik},\l)}-1\Bigr|
=|D_0(e^{ik},\l)|O(|z|^{-1})<|D_0(e^{ik},\l)|
$$
на всех контурах.
По теореме Руше $D(e^{ik},\cdot)$ имеет столько же нулей, сколько
$D_0(e^{ik},\cdot)$ в каждой из ограниченных областей и в оставшейся
неограниченной области. Поскольку
$D_0(e^{ik},\cdot)$ имеет ровно один простой нуль в каждой точке
$(2\pi n+k)^3,n\in\Z$,
и $N'>N$ может быть выбрано произвольно большим,
мы получаем нужный результат.
$\BBox$

\section {Оператор $H( k)$}
\setcounter{equation}{0}

Если $p=q=0$, то каждый оператор $H_0( k)=i\pa^3,
k\in[0,2\pi)$, действующий в $L^2(0,1)$, самосопряжен на области
определения
$$
\mD(H_0( k))=\rt\{f,f'''\in L^2(0,1):f^{(j)}(1)=e^{i k}f^{(j)}(0)\
\all\ j=0,1,2\rt\}.
$$
Собственные значения оператора $H_0( k),k\in[0,2\pi)$, все простые и равны
$\l_{n}^0( k)=(2\pi n+ k)^3,n\in\Z$. Соответствующие
собственные функции $\p_{n, k}^0=e^{i(2\pi n+ k)t}$
образуют ортонормированный базис в $L^2(0,1)$.

\begin{lemma}
\lb{PrA}

i) Каждый оператор $H(k), k\in[0,2\pi)$, -- самосопряженный.

ii) Оператор $H( k),k\in[0,2\pi)$, имеет дискретный спектр
\[
\lb{sAt}
\s(H( k))=\{\l\in\R:e^{i k}\ \text{-- собственное
значение}\  M(1,\l)\}.
\]
Его резольвента $(H(k)-\l)^{-1},\l\in\C\sm\s(H(k))$,
является оператором Гильберта-Шмидта и имеет вид

\[
\lb{res}
\bigl((H(k)-\l)^{-1}f\bigr)(t)=i\int_0^1R_{k,13}(t,s,\l)f(s)ds,\qqq t\in[0,1],
\]
где
\begin{multline}
\lb{e7}
R_ k(t,s,\l)=\bigl(R_{k,jk}(t,s,\l)\bigr)_{j,k=1}^3
\\
= M(t,\l)\Bigl(\chi(t-s)\1_3-\bigl( M(1,\l)-e^{i
k}\1_3\bigr)^{-1}  M(1,\l)\Bigr) M^{-1}(s,\l),
\end{multline}
$\chi(t)=\ca1,\ t\ge 0\\ 0,\ t<0\ac\!\!\!\!$.
Более того, при достаточно большом $\a>0$
резольвента $(H(k)-i\a)^{-1}$ является аналитической операторнозначной функцией
переменной $k$ в некоторой окрестности интервала $[0,2\pi]$ в $\C$.

iii) Собственные функции $\p_{n, k},n\in\Z$, оператора $H( k),k\in[0,2\pi)$,
с собственными значениями $\l_{n}( k)$ образуют ортонормированный базис в $L^2(0,1)$.
Большие по модулю собственные значения $\l_{n}( k)$ -- простые.
Занумеруем $\l_{n}( k)$ в порядке возрастания
$...\le\l_{-1}( k)\le\l_{0}( k)\le\l_{1}( k)\le\l_{2}( k)\le...$
с учетом кратности.
Верна асимптотика
\[
\lb{e9}
\l_{n}( k)=\bigl(2\pi n+ k\bigr)^3\bigl(1+O(n^{-1})\bigr)
\qqq\text{при}\qqq n\to\pm\iy
\]
равномерно по $k$.

\end{lemma}

\no{\bf Доказательство.}
i) Введем следующие операторы в $L^2(0,1)$:
$$
H_+=i\pa^3+i\pa p+ip\pa +q,
$$
$$
\mD(H_+)=\Bigl\{y\in L^2(0,1):i(y''+py)'+ipy'+qy\in
L^2(0,1),y'',(y''+py)'\in L^1([0,1])\Bigr\}
$$
и $H_-$, который является сужением $H_+$ на область
$$
\mD(H_-)=\Bigl\{y\in
\mD(H_+):y(0)=y(1)=y'(0)=y'(1)=(y''+py)(0)=(y''+py)(1)=0\Bigr\}.
$$
Оператор $H_-$ -- плотно определенный симметрический оператор с
индексом дефекта $(3,3)$ и $H_+=H_-^*,H_-=H_+^*$, см. \cite{GM}.
Введем линейные отображения $\G_1,\G_2:\mD(H_+)\to\C^3$ равенствами
\[
\lb{sbv}
\G_1y=\ma i(y''+py)(1)\\-i(y''+py)(0)\\y'(1)-iy'(0)\am,\qq
\G_2y=\ma y(1)\\y(0)\\{2+i\/2}y'(1)-{1+2i\/2}y'(0)\am.
\]
Далее мы используем следующий результат из \cite{GM}:

\no{\it

1) Пусть $A$ -- унитарный оператор в $\C^3$.  Тогда
сужение оператора $H_+$ на множество функций $y$ из $\mD(H_+)$,
удовлетворяющих условию
\[
\lb{e1}
(A-\1_3)\G_1y+i(A+\1_3)\G_2y=0,
\]
является самосопряженным
расширением $H_A$ оператора $H_-$.

2) Для каждого
самосопряженного расширения $\wt H$ оператора $H_-$ существует
унитарный оператор $A$ в $\C^3$ такой, что $\wt H=H_A$.

3) Соотношение
между множеством $\{A\}$ унитарных операторов в $\C^3$ и множеством
$\{h_A\}$ самосопряженных расширений $H_-$ является биекцией. }

Подберем унитарный оператор $A$ так, чтобы условия \er{e1}
для оператора $H_A$ давали условия \er{cDA} для оператора $H(k)$.
Пусть $ k\in[0,2\pi)$ и пусть оператор $A$ имеет вид
\[
\lb{e2}
A=\ma0&-e^{i k}&0\\-e^{-i k}&0&0\\0&0&{ia+b\/ia-b}\am,
\qqq a=ie^{i k}+1,\qq
b={2i-1\/2}e^{i k}+{2-i\/2}.
\]
Равенство
$
({ia+b\/ia-b})^{-1}=\ol{({ia+b\/ia-b})}
$
показывает, что $A$ -- унитарный оператор.
Подставляя \er{e2} и \er{sbv} в равенство \er{e1}, мы получаем
$$
\!\ma-1&-e^{i k}&0\\-e^{-i k}&-1&0\\0&0&{2b\/ia-b}\am\!\!\!
\ma i(y''+py)(1)\\-i(y''+py)(0)\\y'(1)-iy'(0)\am
+\ma 1&-e^{i k}&0\\-e^{-i k}&1&0\\0&0&{i2a\/ia-b}\am\!\!\!
\ma iy(1)\\ iy(0)\\ {2i-1\/2}y'(1)+{2-i\/2}y'(0)\am=0,
$$
что равносильно
\[
\lb{e3}
\ma y\\ y'\\ y''+py\am(1)=e^{i k}\ma y\\ y'\\ y''+py\am(0).
\]
Сравнивая \er{e3} с \er{cDA}, заключаем, что $H_A=H( k)$, где $A$
определен равенством \er{e2}. Поскольку $A$ -- унитарный, $H_A$ и,
следовательно, $H(k)$ -- самосопряженный.

ii)
Решения уравнения $iy'''+(py)'+py'+(q-\l)y=f$ удовлетворяют равенству
\[
\lb{e4} Y(t)= M(t,\l)Y(0) + M(t,\l)\int_0^t M^{-1}(s,\l)F(s)ds,\ \
Y=\ma y\\y'\\y''+py\am,\ \ F=\ma 0\\0\\if\am,
\]
где $(t,\l)\in\R\ts\C$. Условия \er{e3} дают
\[
\lb{e5}
 M(1,\l)Y(0)+ M(1,\l)\int_0^1 M^{-1}(s,\l)F(s) ds=e^{i k}Y(0).
\]
Пусть оператор $( M(1,\l)-e^{i k}\1_3)^{-1}$ ограничен для
некоторого $\l\in\C$. Тогда из равенства \er{e5} имеем
$$
Y(0)=-\bigl( M(1,\l)-e^{i k}\1_3\bigr)^{-1}
 M(1,i\a)\int_0^1 M^{-1}(s,\l)F(s)ds.
$$
Подставляя это равенство в \er{e4}, мы получаем
\[
\lb{e6}
Y(t)=\int_0^1R_ k(t,s,\l)F(s)ds
\qq \text{для всех}\qq  t\in\R,
\]
откуда следует \er{res}.

Матричнозначная
функция $ M$ абсолютно непрерывна по $t$.
Поскольку $\det M(t,\l)=1$, функция $ M^{-1}$ также абсолютно
непрерывна. Тогда равенства \er{res}, \er{e7} показывают, что
резольвента оператора $H( k)$
является оператором Гильберта-Шмидта.
Следовательно, оператор $H( k)$ имеет дискретный спектр.

Пусть $\l$ -- собственное значение оператора
$H(k)$.
Тогда мы имеем
$$
e^{i k}\ma y\\y'\\y''+py\am(0) =\ma y\\y'\\y''+py\am(1) = M(1,\l)\ma
y\\y'\\y''+py\am(0).
$$
Следовательно, $\l$ является нулем функции $D(e^{i k},\l)=\det(
M(1,\l)-e^{i k}\1_3)$. Обратно, пусть $\l$ -- нуль функции $D(e^{i
k},\l)$. Тогда матрица $ M(1,\l)$ имеет собственный вектор
$(x_1,x_2,x_3)^\top$, соответствующий собственному значению $e^{i
k}$. Решение $y$ уравнения \er{1b}, удовлетворяющее начальным
условиям $y^{(j-1)}(0)=x_j,j=1,2,3$, дает собственную функцию
оператора $H( k)$. Таким образом, $\l$ -- собственное значение $H(
k)$, откуда следует \er{sAt}.

Пусть $\l=i\a$ с некоторым достаточно большим $\a>0$.
Тогда $z=e^{i{\pi\/6}}\a^{1\/3},\a^{1\/3}>0$, откуда следует
$$
\Re iz=\Re i\o z=-{\a^{1\/3}\/2},\qqq \Re i\o^2 z=\a^{1\/3}.
$$
Асимптотики \er{awt} показывают, что
$$
\t_j(\l)=e^{-{1\/2}\a^{1\/3}}(1+O(\a^{-{1\/3}})),\qq j=1,2,
\qq \t_3(\l)=e^{\a^{1\/3}}(1+O(\a^{-{1\/3}}))\qq\text{при}
\qq\a\to+\iy.
$$
Таким образом, при достаточно больших $\a>0$ собственные значения
$\t_j(i\a),j=1,2,3$, матрицы $M(1,i\a)$ лежат далеко от единичной
окружности. Тогда оператор $(M(1,i\a)-e^{i k}\1_3)^{-1}$
аналитически зависит от $k$ в некоторой окрестности интервала $[0,2\pi]$.
Из равенства \er{res} следует, что
резольвента $(H(k)-i\a)^{-1}$ является
аналитической операторнозначной функцией
переменной $k$ в этой окрестности.

iii) Резольвента оператора $H(k)$ является компактным
оператором. Как известно (см., напр., \cite{R}, гл. 12.5) в этом
случае собственные функции образуют ортонормированный базис.

Равенство $T=\t_1+\t_2+\t_3$ и асимптотика \er{awt} дают
$$
T(\l)=e^{i\o^2 z}\bigl(1+O(z^{-1})\bigr)\qq\text{при}\qq\l\to\pm\iy,\qq
\ol T(\l)=\bigl(1+O(z^{-1})\bigr)
\ca e^{-i\o z}\ \,\text{при}\ \ \l\to+\iy\\
e^{-iz}\qq\text{при}\ \ \l\to-\iy\ac\!\!\!.
$$
Из равенства \er{cM2} получаем
\[
\lb{D1}
D(e^{i k},\l)
=e^{i2 k}e^{i\o^2 z}\bigl(1+O(z^{-1})\bigr)
-e^{i k}e^{-i\o z}\bigl(1+O(z^{-1})\bigr)
=2ie^{i{3\/2}k+\sqrt3 z}\Bigl(\sin{k-z\/2}+O(z^{-1})\Bigr)
\]
при $\l\to+\iy$ и
\[
\lb{D2}
D(e^{i k},\l)
=e^{i2 k}e^{i\o^2 z}\bigl(1+O(z^{-1})\bigr)
-e^{i k}e^{-iz}\bigl(1+O(z^{-1})\bigr)
=2ie^{i{3\/2}k+\sqrt3|z|}\Bigl(\sin{k+|z|\/2}+O(z^{-1})\Bigr)
\]
при $\l\to-\iy$ равномерно по $k\in[0,2\pi]$.
Множество собственных значений $\l_n(k)$ оператора
$H(k)$ вещественно и совпадает с множеством нулей функции $D(e^{i k},\cdot)$.
Из леммы \ref{CLD} ii) следует, что большие по модулю
нули $D(e^{i k},\cdot)$ -- простые. Кроме того,
$\l_n(k)=(2\pi n+k+\d_n)^3,|\d_n|<{\pi\/2}$
для всех достаточно больших $|n|$.
Подставляя $z=\l_n(k)^{1\/3}=2\pi n+k+\d_n$ в \er{D1}, \er{D2},
мы получаем $\d_n=O(n^{-1})$ при $n\to\pm\iy$, что дает асимптотику \er{e9}.
$\BBox$

\begin{lemma}
\lb{l42}

i) Операторнозначная функция $H(\cdot)$ вещественно аналитична в смысле Като
в некоторой окрестности интервала $[0,2\pi]$ в $\C$.
Кроме того, $H( k)$ и $H(2\pi- k)$ антиунитарно эквивалентны
по отношению к обычному комплексному сопряжению.
В частности, их собственные значения равны и соответствующие собственные
функции комплексно сопряжены.

ii) Для $ k\in(0,2\pi)$ собственные значения оператора
$H( k)$ имеют кратность $1$ или $2$.
Кроме того, для каждого $n\in\Z$ существует конечное число $m_n\ge 0$
значений $ k_\ell\in(0,2\pi),\ell=1,...,m_n,$ таких, что
$\l_n( k_\ell)$ есть собственное значение кратности $2$
оператора $H( k_\ell)$.
Каждая функция $\l_{n}(\cdot),n\in\Z$, непрерывна на $[0,2\pi)$,
аналитична и непостоянна на каждом из интервалов
$(0, k_1),( k_{m_n},2\pi),
( k_\ell, k_{\ell+1}),\ell=1,...,m_n-1$.

iii) Пусть $n\in\Z$. Тогда $L^2(0,1)$-значная функция $\p_{n,k}$
непрерывна по $k$ на $[0,2\pi)$ и
вещественно аналитична по $k$ на каждом из интервалов
$(0, k_1),( k_{m_n},2\pi),
( k_\ell, k_{\ell+1}),\ell=1,...,m_n-1$.

\end{lemma}

\no {\bf Замечание.}
Утверждение ii) леммы о том, что функции
$\l_{n}(\cdot),n\in\Z$, не могут быть постоянными,
является ключевым для доказательства абсолютной
непрерывности спектра оператора $H$.

\no{\bf Доказательство.}
i)
Резольвента $(H(k)-i\a)^{-1}$ при достаточно большом $\a>0$
является аналитической операторнозначной функцией
переменной $k$ в некоторой окрестности интервала $[0,2\pi]$ в $\C$.
Следовательно, операторы $H( k)$
образуют аналитическое семейство в смысле Като в этой окрестности,
см. \cite{RS}, гл.~ XII.2.
Определение \er{cDA} оператора $H( k)$ показывает, что
$H( k)$ и $H(2\pi- k)$ антиунитарно эквивалентны.

ii) Напомним следующий известный результат, см. \cite{Ka},
теорема VII.1.8:

\no {\it Если семейство операторов $A(k)$ аналитично (в смысле Като)
в окрестности нуля, то любая конечная система собственных
значений оператора $A(k)$ представляется ветвями одной
или нескольких аналитических функций, имеющих
самое большее алгебраические особенности в нуле.
}

По лемме \ref{CLD} ii) существует $N\ge 1$, не зависящее от $k$,
такое, что все собственные значения $\l_{n}(k),k\in[0,2\pi),|n|>N$,
-- простые. Отсюда следует, что
каждая функция $\l_{n}(\cdot),|n|>N$, аналитична на $[0,2\pi)$.
Применяя вышеприведенный результат к конечной системе собственных значений
$\l_{n}(k),|n|\le N$, собственных значений оператора
$H(k),k\in[0,2\pi)$, получаем,
что каждая функция $\l_{n}(\cdot),|n|\le N$,
непрерывна и кусочно аналитична на $[0,2\pi)$.

Покажем, что любое собственное значение $\l_n(k),n\in\N$, -- простое
для всех $k\in(0,2\pi)$, за исключением конечного числа значений $k$.
Напомним, что число различных собственных значений
$ M(1,\l)$ равно 3
независимо от $\l$, за исключением некоторых специальных точек
$\l\in\C$. В каждом компактном подмножестве в $\C$ имеется лишь
конечное число таких исключительных точек.
Если $\l$ -- вырожденное
собственное значение $H( k)$, то $e^{i k}$ является
кратным собственным значением матрицы $ M(1,\l)$ и $\l$ -- исключительная точка.
Предположим, что существует бесконечное число значений $ k_\ell\in(0,2\pi)$,
таких, что $\l_n( k_\ell)$ (с некоторым фиксированным $n\in\Z$)
есть вырожденное собственное значение.
Тогда множество $\{ k_\ell,\ell\in\N\}$ имеет по крайней мере
одну предельную точку $ k=\lim_{\ell\to\iy} k_\ell\in[0,2\pi]$
и мы имеем: либо
a) $\lim_{\ell\to\iy}\l_n( k_\ell)=\iy$, либо
b) множество $\{\l_n( k_\ell),\ell\in\N\}$ имеет конечный предел.
Асимптотика \er{e9} показывает, что a) невозможно.
Поскольку в каждом компактном подмножестве в $\C$ имеется
лишь конечное число исключительных значений, b) также невозможно.
Таким образом, существует конечное число $m_n\ge 0$ значений
$ k_\ell\in(0,2\pi)$ таких, что
$\l_n( k_\ell)$ есть вырожденное собственной значение оператора
$H( k_\ell)$. Функция $\l_{n}(\cdot)$ непрерывна на $[0,2\pi)$ и
аналитична на каждом из
интервалов $(0, k_1),( k_\ell, k_{\ell+1}),( k_{m_n},2\pi)$.

Если $\l$ -- собственное значение $H( k)$ кратности 3, то мультипликатор
$\t=e^{i k}$ имеет кратность 3. Равенство
\er{cM} показывает, что в этом случае $\t=1$, откуда следует $ k=0$.

Предположим, что  для некоторого $n\in\Z$
функция $\l_n( k)=c=\const$ на некотором непустом
интервале $(\a,\b)\ss[0,2\pi)$, и пусть
$A=T(c)$. Тогда из \er{cM2} следует равенство
\[
\lb{kcon}
-e^{i3 k}+e^{i2 k}A-e^{i k}\ol A+1=0\qq\all\qq k\in(\a,\b).
\]
Выражая отсюда $e^{ik}$ по формулам
Кардано, мы получаем $k=\const$, то есть равенство \er{kcon}
не может выполняться для всех $ k\in(\a,\b)$. Полученное противоречие
доказывает утверждение.

iii) Результат следует из i), ii) и теоремы Като-Реллиха,
см.~\cite{RS}, теорема~XII.8.
$\BBox$

\section {Спектр оператора $H$}
\setcounter{equation}{0}

Обозначим через $C_0^\iy(\R)$ (плотное в $L^2(\R)$) множество гладких
функций на $\R$ с ограниченным носителем.

\begin{lemma}
\lb{uU}
i) Продолжим $\p_{n, k},(n, k)\in\Z\ts[0,2\pi)$, с $[0,1]$ на $\R$
равенством $\p_{n, k}(t+1)=e^{i k}\p_{n, k}(t)$.
Для $f\in C_0^\iy(\R)$ введем
\[
\lb{dwt}
\wt f_n( k)=\int_\R f(t)\ol{\p_{n, k}(t)}dt.
\]
Верны равенства
\[
\lb{PF}
\|f\|^2={1\/2\pi}\sum_{n\in\Z}\int_0^{2\pi}|\wt f_n( k)|^2d k,
\qq
f(t)={1\/2\pi}\sum_{n\in\Z}\int_0^{2\pi}\wt f_n( k)\p_{n, k}(t)d k,\qq t\in\R.
\]

ii) Оператор
$H_1=U^{-1}\int_{[0,2\pi)}^\os H( k){d k\/2\pi}\ U$
-- самосопряженный на области
$\mD(H_1)=\mD(H)$, где область $\mD(H)$ определена в \er{cDH}.
Кроме того,
оператор $H_1$ удовлетворяет
равенству
\[
\lb{mH}
(\wt{H_1 f})_n( k)=\l_n( k)\wt f_n( k)\qq\text{для всех}\qq
(n, k)\in\Z\ts[0,2\pi),\qq f\in\mD(H_1),
\]
где операция $\wt{\ }$ считается продолженной на все $L^2(\R)$ по непрерывности.

\end{lemma}

\no {\bf Доказательство.} Напомним, что унитарный  оператор
$U:L^2(\R)\to\mH$ действует по формуле
\[
\lb{gt5} u_k(t)=(U f)_k(t)=\sum_{\ell\in\Z}e^{-i\ell k}f(t+\ell).
\]
Здесь $\mH=\int_{[0,2\pi)}^\os \mH'\, {d k\/2\pi},\ \mH'=L^2([0,1],dt)$ и введем
скалярное произведение $(\cdot,\cdot)_0$ и норму
$\|\cdot\|_0$ в гильбертовом пространстве $\mH'$.

i) Для $f\in C_0^\iy(\R)$ сумма в \er{gt5}
конечна и $u_ k$ удовлетворяет равенству
\[
\lb{et5}
u_{ k}(t+1)=e^{-i k}u_{ k}(t),\qq
f(t)={1\/2\pi}\int_0^{2\pi}u_ k(t)d k\qq  \all \qq t\in\R.
\]
Кроме того,
$$
\wt f_n( k)=\int_0^1\sum_{\ell\in\Z} f(t+\ell)\ol{\p_{n, k}(t+\ell)}dt
=\int_0^1\sum_{\ell\in\Z} e^{-ik \ell}f(t+\ell)\ol{\p_{n, k}(t)}dt
=\int_0^1u_ k(t)\ol{\p_{n, k}(t)}dt.
$$
Поскольку $\p_{n, k}$ образуют ортонормированный базис в $L^2(0,1)$, мы имеем
\[
\lb{et2}
\sum_{n\in\Z}|\wt f_n( k)|^2=\int_0^1|u_ k(t)|^2dt,
\qq
u_ k(t)=\sum_{n\in\Z}\wt f_n( k)\p_{n, k}(t)
\qq\all\qq t\in[0,1].
\]
Унитарность $U$ дает $ \|f\|^2=\|u\|_\mH^2={1\/2\pi}\int_0^{2\pi}d
k\int_0^1|u_ k(t)|^2dt. $ Из этого равенства и первого равенства в
\er{et2} следует первое равенство в \er{PF}. Подставляя второе
равенство в \er{et2} во второе равенство в \er{et5}, получаем второе
равенство в \er{PF}.

ii) Оператор $A=\int_{[0,2\pi)}^\os H( k){d k\/2\pi}$ на функциях из
области определения
$$
\mD(A)=\{u\in\mH:u_ k\in\mD(H( k))\ \all\  k\in[0,2\pi),\
\int_0^{2\pi}\|H( k)u_ k\|_0^2d k<\iy\}
$$
-- самосопряженный, см. \cite{RS}, теорема XIII.85. Тогда
$H_1=U^{-1}AU$ самосопряжен
на области $\mD(H_1)=U^{-1}\mD(A)$.

Покажем, что $\mD(H_1)=\mD(H)$, где $\mD(H)$ -- область, заданная
равенством \er{cDH}. Пусть $f\in\mD(H)$. Из равенства \er{gt}
следует, что $u_ k=(Uf)( k)\in\mD(H( k))$ для всех $ k\in[0,2\pi)$.
Кроме того, мы имеем $hu_ k=(U(hf))( k)$, где $h=i(\pa^2+p)\pa+i \pa
p+q$. Тогда ${1\/2\pi}\int_0^{2\pi}\|H( k)u_ k\|_0^2d k =\|hf\|^2$.
Из $hf\in L^2(\R)$ следует $\int_0^{2\pi}\|H( k)u_ k\|_0^2d k<\iy$.
Таким образом, $u=Uf\in\mD(A)$ и, следовательно, $U\mD(H)\ss\mD(A)$,
откуда имеем $\mD(H)\ss U^{-1}\mD(A)=\mD(H_1)$. Обратно, пусть
$u\in\mD(A)$. Тогда $u_ k\in\mD(H( k))$ для всех $ k\in[0,2\pi)$ и
равенство $f(t)={1\/2\pi}\int_0^{2\pi}u_ k(t)d k$ дает $f',f''+pf\in
AC(\R)$. Кроме того, $\int_0^{2\pi}\|H( k)u_ k\|_0^2d k<\iy$.
Используя унитарность $U$, отсюда получаем $ \|hf\|^2
={1\/2\pi}\int_0^{2\pi}\|H( k)u_ k\|_0^2d k<\iy, $ что означает
$hf\in L^2(\R)$. Таким образом $f=U^{-1}A\in\mD(H)$ и мы имеем
$U^{-1}\mD(A)=\mD(H_1)\ss\mD(H)$. Следовательно, $\mD(H_1)=\mD(H)$.

Докажем \er{mH}.
Пусть $f\in\mD(H_1)\cap C_0^\iy(\R)$. Известно, что множество $\mD(H_1)\cap C_0^\iy(\R)$
плотно в $L^2(\R)$ (см., напр., \cite{EM}, Appendix A).
Равенства \er{dwt}, \er{gt} дают
\[
\lb{e10} \wt{(H_1f)}_n( k)=\int_\R (H_1f)(t)\ol{\p_{n, k}(t)}dt
=\sum_{\ell\in\Z}\int_0^1 (H_1f)(t+\ell)\ol{\p_{n, k}(t)}e^{-i\ell
k}dt =((UH_1f)(k),\p_{n, k})_{0}
\]
и это равенство по непрерывности может быть продолжено на все множество
$f\in\mD(H_1)$.
Из определения $H_1$ следует $(UH_1f)(k)=(AUf)(k)=H(k)u_ k$.
Подставляя это равенство в \er{e10}, мы получаем
$$
\wt{(H_1f)}_n( k)=(H(k)u_ k,\p_{n, k})_{0} =\l_n( k)(u_k,\p_{n,
k})_{0},\qq f\in\mD(H_1).
$$
Из второго равенства \er{et2} имеем $(u_k,\p_{n, k})_{0}=\wt
f_n( k)$, откуда следует \er{mH}. $\BBox$

\no {\bf Замечания.}
1) В предложении \ref{thH} мы доказываем, что $H_1=H$.
Функции $\p_{n, k},(n, k)\in\Z\ts[0,2\pi)$, продолженные
на все $t\in\R$ равенством $\p_{n, k}(t+1)=e^{i k}\p_{n, k}(t)$,
являются решениями Флоке для оператора $H$.
Равенства \er{PF} дают разложение функции $f$ в интеграл решений Флоке
и равенство Парсеваля для этого разложения.
Равенство \er{mH} показывает существование разложения по собственным
функциям для $H$.

2) Результаты, аналогичные результатам леммы \ref{uU}, хорошо известны
для оператора второго порядка, см., напр., \cite{RS}, гл. XIII.16.

\no {\bf Доказательство предложения \ref{thH}.}
Пусть $f\in\mD(H_1)=\mD(H)$.
Интегрирование по частям дает
\[
\lb{fH}
(\wt{H f})_n( k)=\int_\R (Hf)(t)\ol{\p_{n, k}(t)}dt
=\int_\R f(t)\bigl((i\pa^3+i p\pa+i \pa p+q)\ol\p_{n, k}\bigr)(t)dt
=\l_n( k)\wt f_n( k)
\]
для всех $(n, k)\in\Z\ts[0,2\pi).$
Равенства \er{mH}, \er{fH} дают
$(\wt{(H-H_1) f})_n( k)=0$ для всех $(n, k)\in\Z\ts[0,2\pi)$.
Тогда из равенства \er{PF} имеем
$(H-H_1) f=0$ для всех $f\in\mD(H)$, следовательно, $H=H_1$.
Из леммы \ref{uU} ii)
следует самосопряженность $H$ и равенство \er{deH}.
$\BBox$

В теореме \ref{spec} мы, наряду с \er{sdD}, доказываем равенство
\[
\lb{spl}
\s(H)=\cup_{n\in\Z}\l_n([0,2\pi)).
\]

\no {\bf Доказательство теоремы \ref{spec}.}
i) Мы используем следующие известные результаты,
см. \cite{RS}, теоремы XIII.85,86.

{\it Пусть $A=\int_{[0,2\pi)}H( k){d k\/2\pi}$ и для каждого $ k\in[0,2\pi)$
$H( k)$ -- самосопряженные операторы в $L^2(0,1)$.
Предположим, что нам даны $L^2(0,1)$-значные функции $\{\p_{n}(\cdot)\}_{n\in\Z}$
на $[0,2\pi]$, непрерывные на $[0,2\pi)$, кусочно вещественно аналитические на $(0,2\pi)$,
и комплекснозначные функции $\l_n(\cdot)$, непрерывные на $[0,2\pi)$,
кусочно аналитические на $[0,2\pi)$, такие, что:

a) каждая функция $\l_n(\cdot)$ непостоянна
на любом подинтервале в $[0,2\pi)$;

b) $H( k)\p_n( k)=\l_n( k)\p_n( k)$
для всех $(n, k)\in\Z\ts[0,2\pi)$;

c) для каждого $ k\in[0,2\pi)$ набор $\{\p_{n}( k)\}_{n\in\Z}$
образует ортонормированный базис в $L^2(0,1)$.

\no Тогда $A$ имеет чисто абсолютно непрерывный спектр и
$\s(A)=\ol{\cup_{ k\in[0,2\pi)}\s(H( k))}$.

}

Из этих результатов и лемм \ref{PrA}, \ref{l42} следует, что
спектр оператора $H$ чисто абсолютно непрерывен и
удовлетворяет \er{spl} и первому равенству в \er{sdD}.
Равенства $\D_j={1\/2}(\t_j+\t_j^{-1}), j=1,2,3$, дают второе равенство в \er{sdD}.

ii) Теорема \ref{TMM} ii) показывает, что $\s(H)=\R$ и
для любого $\l\in\s(H)$ имеется только две возможности:
a) ровно один мультипликатор  $\t(\l)$ лежит на единичной окружности;
b) все три мультипликатора лежат на единичной окружности.

Рассмотрим случай a): ровно один мультипликатор  $\t(\l)$ лежит на единичной окружности
для всех $\l\in(\a,\b)$ с некоторыми $\a<\b$. Тогда $(\a,\b)\in\s(H)$
и равенство \er{mH} показывает, что
спектральный проектор $\chi_{(\a,\b)}(H)$
унитарно эквивалентен оператору умножения на $\l\in(\a,\b)$.
Поэтому спектр $H$ в интервале $(\a,\b)$ имеет кратность 1.

Рассмотрим случай b): все три мультипликатора лежат на единичной окружности
для всех $\l\in(\a,\b)$ с некоторыми $\a<\b$. Тогда
спектральный проектор $\chi_{(\a,\b)}(H)$
унитарно эквивалентен оператору умножения на $\l\1_3,\l\in(\a,\b)$.
Следовательно, спектр $H$ в интервале $(\a,\b)$ имеет кратность 3.

Асимптотика \er{awt} показывает, что
ровно один мультипликатор удовлетворяет условию \er{sdD}
при $\l\in\R\sm[-R,R]$  с достаточно большим $R>0$,
значит спектр имеет кратность 1 при таких $\l$.

Пусть $\t_j=e^{ik_j}$  для всех $j=1,2,3$.
По теореме \ref{T1r} i) все $k_j(\l),j=1,2,3$, различны при всех $\l\in\C$,
кроме некоторых исключительных значений $\l$, и число таких исключительных
значений конечно в каждой конечной области.
Равенства \er{r} и $e^{i(k_1+k_2+k_3)}=1$ дают
\[
\lb{rqm}
\r=(e^{ik_1}-e^{ik_2})^2
(e^{ik_1}-e^{ik_3})^2
(e^{ik_2}-e^{ik_3})^2
=-64\sin^2{k_1-k_2\/2}
\sin^2{k_1-k_3\/2}
\sin^2{k_2-k_3\/2}.
\]
Если спектр в точке $\l$ имеет кратность 3, то, по теореме \ref{TMM} iii),
$k_j(\l)\in\R$ для всех $j=1,2,3$, и \er{rqm} дает $\r(\l)\le 0$.
Если спектр в точке $\l$ имеет кратность 1, то ровно одно значение $k_j$,
скажем, $k_1$, вещественно и $k_2=\ol k_3$ - невещественны.
Из равенства \er{rqm} следует $\r(\l)>0$, откуда получаем \er{sro}.

iii) Доказательство утверждения стандартное, приводим его для полноты.
Пусть $\l_0\in\R$ удовлетворяет условиям: $\D_j(\l_0)\in(-1,1)$
для некоторого $j=1,2,3$,
$\l_0$ не является точкой ветвления функции $\D_j$ и $\D_j'(\l_0)=0$.
Тогда $\D_j(\l)=\D_j(\l_0)+{1\/2}\D_j''(\l_0)(\l-\l_0)^2+O((\l-\l_0)^3)$
при $\l-\l_0\to 0$.
Рассмотрим отображение $\l\to \D_j(\l)$ в некоторой окрестности точки
$\l_0$. Любой угол, образованный линиями, начинающимися в точке $\l_0$,
переводится этим отображением в угол, в $2$ или более раз больший. Тогда отрезок
$[\D_j(\l_0)-\d,\D_j(\l_0)+\d]\ss[-1,1]$ для некоторого достаточно малого $\d>0$
имеет прообраз, который не может целиком лежать на вещественной оси.
Из равенства \er{sdD} следует, что $H$ имеет невещественный спектр,
что противоречит самосопряженности.
Таким образом, $\D_j'(\l_0)\ne 0$, что и доказывает утверждение.
$\BBox$

Работа выполнена при финансовой поддержке Министерства образования и
науки Российской Федерации, ГК № 14.740.11.0581.

\end{document}